\documentclass[cmp]{svjour}
\usepackage{graphics}
\usepackage{amsmath}
 \usepackage{epsfig}
\usepackage{amssymb}
\usepackage{amsfonts}
 \usepackage{latexsym}

\begin{document}

\topmargin=2cm

\newtheorem{Proposition}{Proposition}
\newtheorem{Comment}{Proposition}
  \newtheorem{Remark}[Proposition]{Remark}
  \newtheorem{Corollary}[Proposition]{Corollary}
  \newtheorem{Lemma}[Proposition]{Lemma}
    \newtheorem{Theorem}[Proposition]{Theorem}
  \newtheorem{Note}[Proposition]{Note}
\newtheorem{Definition}{Definition}
\def\bfy{\mathbf{y}}
\def\bfz{\mathbf{z}}
\def\bfC{\mathbf{C}}
\def\Nsm{\hbox{\small I\hskip -2pt N}}
\def\rdb{\hbox{ I\hskip -2pt R}}
\def\cdb{\hbox{\it l\hskip -5.5pt C\/}}
\def\ndd{\hbox{\it I\hskip -2pt N}}
\def\zdd{\hbox{\sf Z\hskip -4pt Z}}
\def\e{\epsilon}
\def\ei{\epsilon^{-1}}
\def\a{\alpha}
\def\l{\lambda}
\def\tl{\tilde\lambda}
\def\ct{{\mbox{const}}}
\def\d{\delta}
\def\cf{{\cal F}}
    \def\z{\noindent}  
 \def\Box{{\hfill\hbox{\enspace${\sqre}$}} \smallskip}
    \def\sqr#1#2{{\vcenter{\vbox{\hrule height .#2pt
                             \hbox{\vrule width .#2pt height#1pt \kern#1pt
                                   \vrule width .#2pt}
                             \hrule height .#2pt}}}}
 \def\sqre{\mathchoice\sqr54\sqr54\sqr{4.1}3\sqr{3.5}3}     
     \def\erm{\mathrm{e}}
    \def\irm{\mathrm{i}}
    \def\drm{\mathrm{d}}
 \def\bchi{\mbox{\raisebox{.4ex}{\begin{Large}$\chi$\end{Large}}}}
     \def\CC{\mathbb{C}}
    \def\DD{\mathbb{D}}
    \def\NN{\mathbb{N}}
    \def\QQ{\mathbb{Q}}
    \def\RR{\mathbb{R}}
    \def\ZZ{\mathbb{Z}}

\title{Evolution of a model quantum system under time periodic
forcing: conditions for complete ionization}
\titlerunning{Ionization of simple model}

%\subtitle{ aa}
\author{O. Costin\inst{1}\and R. D. Costin\inst{1} \and
  J. L. Lebowitz\inst{1}\and A. Rokhlenko\inst{1}% etc
}                     % Do not remove
\institute{ Department of Mathematics\\ Rutgers University
\\ Piscataway, NJ 08854-8019}
\date{Received: date / Revised version: date}
\maketitle

 \today

\vskip 0.5cm 

\begin{abstract}
  We analyze the time evolution of a one-dimensional quantum system with
  an attractive delta function potential whose strength is subjected to
  a time periodic (zero mean) parametric variation $\eta(t)$.  We show
  that for generic $\eta(t)$, which includes the sum of any finite
  number of harmonics, the system, started in a bound state will get
  fully ionized as $t\rightarrow\infty$. This is irrespective of the
  magnitude or frequency (resonant or not) of $\eta(t)$.  There are
  however exceptional, very non-generic $\eta(t)$, that do not lead
  to full ionization, which include rather simple explicit periodic
  functions. For these $\eta(t)$ the system evolves to a nontrivial
  localized stationary state which is related to eigenfunctions of the
  Floquet operator.
%end of abstract
\end{abstract}

\section{Introduction and  results}
\label{intro}

We are interested in the qualitative long time behavior of a quantum system evolving under a time
dependent Hamiltonian
$H(t)=H_0+H_1(t)$, i.e.\ in 
the nature of
the solutions of the Schr{\"o}dinger equation 

\begin{equation}
  \label{eq:Sc1}
  i\hbar\partial_t\psi=[H_0+H_1(t)]\psi
\end{equation}
Here $\psi$ is the wavefunction of the system, belonging to some Hilbert
space ${\cal H}$, $H_0$ and $H_1$ are Hermitian operators and
equation(1) is to be solved subject to some initial condition $\psi_0$.
Such questions about the solutions of (\ref{eq:Sc1}) belong to what
Simon \cite{[1]} calls ``second level foundation'' problems of quantum
mechanics. They are  of particular practical interest for the
ionization of atoms and/or dissociation of molecules, in the case when
$H_0$ has both a discrete and a continuous spectrum corresponding
respectively to spatially localized (bound) and scattering (free) states
in $\RR^d$. Starting at time zero with the system in a bound state and
then ``switching on'' at $t=0$ an external potential $H_1(t)$, we want
to know the ``probability of survival'', $P(t)$, of the bound states, at
times $t>0$: $P(t)=\sum_j|\langle \psi(t),u_j\rangle |^2$, where the sum
is over all the bound states $u_j$ \cite{[2],[3],[4],[5],[6],[7],[8]}.

This problem has been investigated both analytically and numerically for
the case $H_1(t)=\eta(t)V_1(x)$ with $\eta(t)=r\sin(\omega t+\theta)$
and $V_1$ a time independent potential, $x \in \RR^d$. When $\omega$ is
sufficiently large for ``one photon'' ionization to take place, i.e.,
when $\hbar\omega>-E_0$, $E_0$ the energy of the bound (e.g.\ ground)
state of $H_0$ and $r$ is ``small enough'' for $H_1$ to be treated as a
perturbation of $H_0$ then this is a problem discussed extensively in
the literature (\cite{[7],[8]}). Starting with the system in its ground
state the long time behavior of $P(t)$ is there asserted to be given by
the $P(t)\sim \exp[-\Gamma_F t]$. The rate constant $\Gamma_F$ is
computed from first order perturbation theory according to Fermi's
golden rule.  It is proportional to the square of the matrix element
between the bound and free states, multiplied by the appropriate density
of continuum states in the vicinity of the final state which will have
energy $\hbar\omega-E_0$ \cite{[6],[7],[8],[9]}.

Going from perturbation theory to an exponential decay involves
heuristics based on deep physical insights requiring assumptions which
seem very hard to prove.  It is therefore very gratifying that many
features of this scenario have been recently made mathematically
rigorous by Soffer and Weinstein \cite{[6]} (their analysis was
generalized by Soffer and Costin \cite{[6a]}). They considered the case
when $H_0=-\nabla^2+V_0(x)$, $x\in\RR^3$, $V_0$ compactly supported and
such that there is exactly one bound state with energy $-\omega_0$ (from
now on we use units in which $\hbar=2m=1$) and a continuum of
quasi-energy states with energies $k^2$ for all $k\in\RR^3$. The
perturbing potential is $H_1(t)=r\cos(\omega t)V_1(x)$ with $V_1(x)$
also of compact support and satisfying some technical conditions.  They
then showed that for $\omega>\omega_0$ and $r$ small enough there is
indeed an intermediate time regime where $P(t)$ has a dominant
exponential form with the Fermi exponent $\Gamma_F$. This regime is
followed for longer times by an inverse power law decay.  Some of these
restrictions can presumably be relaxed but the requirement that $r$ be
small is crucial to their method which is essentially perturbative.

The behavior of $P(t)$ becomes much more difficult to analyze when the
strength of $H_1(t)$ is not small and perturbation theory is no longer a
useful guide. This became clear in the seventies with the beautiful
experiments by Bayfield and Koch, c.f.\ \cite{[10]} for a review, on the
ionization of highly excited Rydberg (e.g.\ hydrogen atoms) by intense
microwave electric fields. These experiments, showed quite unexpected
nonlinear behavior of $P(t)$ as a function of the initial state, field
strength $E$ and the frequency $\omega$. These results as well as other
multiphoton ionizations of hydrogen atoms have been (and continue to be)
analyzed by various authors using a variety of methods.  Prominent among
these are semi-classical phase-space analysis, numerical integration of
the Schr\"odinger equation, Floquet theory, complex dilation, etc. While
the results obtained so far are not rigorous, they do give physical
insights and quite good agreement with experiments although many
questions still remain open even on the physical level
\cite{[10],[11],[12],[13],[14]}.

In addition to the above experiments on Rydberg atoms there are also
many  experiments which use strong laser fields to produce
multiphoton ($\omega<-E_0$) ionization of multielectron atoms
and/or dissociation of molecules \cite{[15],[16]}. These systems are more
complex than Rydberg atoms and their analysis is correspondingly less
developed.  One unexpected result of certain studies is that  an
increase in the intensity of the field may reduce the degree of
ionization, i.e., $P(t)$ can be non monotone in the field strength $E$ at
large values of $E$. This phenomenon, which is often called
``stabilization'', can be observed in some numerical simulations,
analyzed rigorously in some models and is claimed to have been
seen experimentally c.f. \cite{[5]} and \cite{[17],[18],[19],[20]}.

It turns out that many features observed for Rydberg atoms and also
stabilization are already present in a simple model system which we have
recently begun to investigate analytically \cite{[21],[22],[23]}.  
The results described there show that  phenomenon of ionization by periodic fields is 
very complex indeed once one goes beyond the perturbative regime.  This will
become even clearer from the new results about this model presented here.

\section{The model}
\label{model}

We consider a very simple quantum system where we can analyze rigorously
many of the phenomena expected to occur in more realistic systems
described by (\ref{eq:Sc1}).  This is a one dimensional system with an
attractive delta function potential.  The unperturbed Hamiltonian $H_0$
has, in suitable units, the form

\begin{equation}
  \label{eq:(1)}
  H_0=-{\frac{\mathrm{d}^2}{\mathrm{d}x^2}}-
2\,\delta
  (x),\ \ -\infty<x<\infty.
\end{equation}
The zero range (delta--function) attractive potential is much used in
the literature to model short range attractive potentials
\cite{[24],[25],[26],[27]}.  It belongs, in one dimension, to the class $K_1$
\cite{[2]}.  $H_0$ has a single bound state $ u_b(x)=e^{-|x|}$ with
energy $ -\omega_0=-1$. It also has continuous uniform spectrum on the
positive real line, with generalized eigenfunctions
$$u(k,x)=\frac{1}{\sqrt{2\pi}}\left
(e^{ikx}-\frac{1}{1+i|k|}e^{i|kx|} \right ), \ \ -\infty<k<\infty$$

\z and energies $k^2$.  

Beginning at  $t=0$, we apply a parametric perturbing
potential, i.e. for $t>0$ we have

\begin{equation}
  \label{eq:timedep}
  H(t)=H_0 -2\, \eta(t)\delta(x)
\end{equation}
\z and  solve the time  dependent Schr{\"o}dinger
equation (\ref{eq:Sc1})
 for $\psi(x,t)$, with $\psi(x,0) = \psi_0(x)$.  Expanding $\psi$ in
eigenstates of $H_0$ we write
\begin{eqnarray}
  \label{eq:(2)}
 \psi (x,t)=\theta (t)u_b(x)e^{it}\hskip 4cm\nonumber\\ \hskip
1cm +\int_{-\infty}^
{\infty}\Theta (k,t)u(k,x)e^{-i k^2 t}dk \ \ (t\geq 0)
\end{eqnarray}
with initial values $\theta (0)=\theta_0,\ \Theta (k,0)=\Theta_0(k)$  suitably normalized, 

\begin{equation}
  \label{eq:e4}
\langle \psi_0, \psi_0 \rangle =  |\theta_0|^2+\int_{-\infty}^\infty|\Theta_0(k)|^2dk=1
\end{equation}

We then have that the survival probability of the bound state is  $P(t) =
|\theta(t)|^2$, while  $|\Theta(k,t)|^2 dk$ gives the  
``fraction of ejected particles'' with 
(quasi-) momentum in the interval $dk$.

This problem can be reduced
to the solution of an integral equation in a single variable \cite{[21],[22]}.   Setting

\begin{equation}
  \label{eq:eqY}
  Y(t)=\psi(x=0,t)\eta(t)e^{it}
\end{equation}

\z we have

\begin{eqnarray}
  \label{eq:(3)}
  &\theta (t)=\theta_0+2i\int_0^t Y(s) ds \ , \\
  &\Theta(k,t)= \Theta_0(k)+2|k|/\big[\sqrt{2\pi} (1-i|k|)\big]\int_0^t Y(s)
e^{i(1+k^2)s} ds \ .
\end{eqnarray}

\z  $Y(t)$ satisfies the integral equation

\begin{equation}
  \label{eq:(5)}
  Y(t)=\eta(t)\left
\{I(t)+\int_0^t [2i+M(t-t')]Y(t') dt'\right \}=\eta(t)\Big(I(t)+(2i+M)*Y\Big)
\end{equation}

\z where the inhomogeneous term is

$$I(t)=\theta_0+\frac{i}{\sqrt{2\pi}}\int_0^{\infty}\frac{\Theta_0(k)+\Theta_0(-k)}{1+ik}e^{-i(k^2+1)t}dk,$$

\z and

$$
M(s)=\frac{2i}{\pi}\int_0^\infty \frac{u^2e^{-is(1+u^2)}}{1+u^2}du=
\frac{1+i}{2\sqrt{2}\pi}\int_s^\infty\frac{e^{-iu}}{u^{3/2}} du
$$ 

\z with

$$f*g=\int_0^tf(s)g(t-s)ds$$

\z In our previous works we considered the case where $\Theta_0(k)
= 0$ and $\eta(t)$ is a finite sum of harmonics with period $2\pi
\omega^{-1}$.  In particular, we showed in \cite{[22]} how to compute the survival probability $P(t)$ as a function of the strength
$r$ and frequency $\omega$ when $\eta(t)=r \sin \omega t$.  Here we study
the general periodic case and write
\renewcommand{\labelenumi}{(\alph{enumi})}

$$\eta=\sum_{j=0}^{\infty}\Big(C_j e^{i\omega j t}+C_{-j}e^{-i\omega j
t}\Big)$$

\z Our assumptions on the $C_j$ are

\begin{enumerate}

\item{$0\not\equiv\eta\in L^{\infty}(\mathbb{T})$}

\item{$C_0=0$}

\item{$C_{-j}=\overline{C_j}$}

\end{enumerate}
{\bf Genericity condition (g)}. Consider the right shift operator ${T}$
on $l_2(\NN)$ given by

$${T}(C_1,C_2,...,C_n,...)=(C_2,C_3,...,C_{n+1},...)$$

We say that ${\bf  C}\in l_2(\NN)$ is {\em generic with respect to
${T}$}
if the Hilbert space generated by all the translates of ${\bf C}$ contains
the vector $e_1=(1,0,0...,)$ (which is the kernel of ${T}$):

\begin{equation}
  \label{eq:cyclic}
e_1\in\bigvee_{n=0}^{\infty}{T}^n \bf C
\end{equation}

\z (where the right side of (\ref{eq:cyclic}) denotes the closure of the
space generated by the ${T}^n \bf C$ with $n\geq 0$.) This condition is
generically satisfied, and is obviously weaker than the ``cyclicity''
condition $ l_2(\NN)\ominus \bigvee_{n=0}^{\infty}{T}^n \bf C=\{0\}$,
which is also generic \cite{Nikol'skii} (Appendix B discusses in more
detail the rather subtle cyclicity condition).

An important case, which satisfies (\ref{eq:cyclic}), (but fails the
cyclicity condition) corresponds to $\eta$ being a trigonometric
polynomial, namely $\bf C\not\equiv 0$ but $C_n=0$ for all large enough
$n$. (We can in fact replace $\mathbf{e}_1$ in (\ref{eq:cyclic}) by
$\mathbf{e}_k$ with any $k\ge 1$.)  A simple example which fails
(\ref{eq:cyclic}) is

\begin{equation}
  \label{eq:counterex}
  \eta(t)=2r\lambda\frac{\lambda-\cos(\omega
    t)}{1+\lambda^2-2\lambda\cos(\omega t)}
\end{equation}
\z for some $\lambda\in (0,1)$, for which $C_n=-r\lambda^n$ for $n\geq
1$. In this case the space generated by $T^n\bf C$ is
one-dimensional. We will prove that there are values of $r$ and
$\lambda$ for which the ionization is incomplete, i.e.\ $\theta(t)$ does
not go to zero for large $t$.

\section{Results and Remarks}

\begin{Theorem}\label{T1}
  Under assumptions (a)...(c) and (g), the survival probability $P(t)$ of
the bound state $u_b$, 
  $|\theta(t)|^2$ tends to zero as $t\rightarrow\infty$.
\end{Theorem}

\begin{Theorem}\label{T2}
  For $\psi_0(x)=u_b(x)$  there exist values of
  $\lambda$, $\omega$ and $r$ in (\ref{eq:counterex}), for which
  $|\theta(t)|\not\rightarrow 0$ as
  $t\rightarrow\infty$.
\end{Theorem}

\z {\bf Remarks}.

1. Theorem~\ref{T1} can be extended to show that $\int_D|\psi(x,t)|^2dx
\rightarrow 0$ for any compact interval $D\in\RR$. This means that the
initially localized particle really wanders off to infinity since by
unitarity of the evolution $\int_\RR|\psi(x,t)|^2dx=1$.
Theorem~\ref{T2} can be extended to show that for some fixed $r$ and
$\omega$ in (\ref{eq:counterex}) there are infinitely many $\lambda$,
accumulating at $1$, for which $\theta(t)\not\rightarrow 0$. In these
cases, it can also be shown that for large $t$, $\theta$ approaches a
quasiperiodic function.

2.  While Theorem~\ref{T1}  holds 
for arbitrary $\psi_0$, care has to be taken with the initial conditions
for Theorem~\ref{T2}.  In particular we cannot have an initial state
such that
in (\ref{eq:(5)})
$I(t)=0$
for all $t$. This would occur, for example, if 
$\psi_0(x)$
is an odd function of $x$.  In that case the evolution takes place
as if the particle was entirely free --- never feeling the delta
function potential. There may also be other special $\psi_0$ for which
$\theta_0 \ne 0$ but for which $\theta(t) \to 0$ as $t \to \infty$.
We have therefore stated Theorem~\ref{T2} for the case $\psi_0 = u_b$.  
We shall also, for simplicity, use this choice of $\psi_0$ in the
proofs of Theorem~\ref{T1}.
For this 
case, which is natural from the physical point of view,
$I(t)=1$ in   (\ref{eq:(5)}). The extension to general $\psi_0$ is
immediate and is given at the end of Section \ref{S5}.

3. In \cite{[22]} we gave a detailed picture of how the decay of
$\theta(t)$ depends on $r$ and $\omega$ when $\eta(t)=r\sin(\omega t)$,
$\theta_0=1$. For small $r$ and $\omega^{-1}$ not too close to an
integer we get an exponential decay with a decay rate
$\Gamma(r,\omega)\sim r^{2(1+\lfloor \omega^{-1}\rfloor)}$ where
$\lfloor\omega^{-1} \rfloor$ is the integer part of $\omega^{-1}$. (For
$\omega > 1$, this corresponds to $\Gamma \sim \Gamma_F$).  At times
large compared to $\Gamma^{-1}$, $|\theta(t)|$ decays as $t^{-3/2}$.
The picture becomes much more complicated when $r$ is large and/or
$\omega^{-1}$ is an integer. In particular there is no monotonicity in
$|\theta(t)|$ as a function of $r$. In \cite{[23]} we proved complete
ionization for the case where $C_n=0$ for $n>N$, $N \geq 1$.

4. We note here that Pillet \cite{[3]} proved complete ionization for
quite general $H_0$ under the assumption that $H_1(t)$ is ``very
random'', in fact a Markov process. Our results are not only consistent
with this but support the expectation that generic perturbations will
lead to complete ionization for general $H_0$. This is what we expect
from entropic considerations --- there is just too much phase space
``out there''. The surprising thing is that even for our simple example
one can readily find exceptions to the rule.

We should also mention here the work of Martin et al. \cite{Martin,Martin2} who
consider the case where $H_0$ has an isolated eigenvalue $E_0$ plus an
absolutely continuous spectrum in the interval $[0,E_{max}]$.  They show
that if the frequency $\omega$ of the periodic, small, perturbation
$H_1(t)$ is larger than $E_0$ then the bound state is stable.  This can
be understood in terms of Fermi's golden rule by noting that the density
of states at the energy $E_0+\omega>E_{max}$ is zero so that $\Gamma_F$
would be zero.

5. There is a direct connection between our results and Floquet theory
where, for a time-periodic Hamiltonian $H(t)$ with period
$T=2\pi/\omega$, one constructs a quasienergy operator (QEO)
\cite{[2],[30],[31]}.

$$K=-i\frac{\partial}{\partial\theta}+H(\theta)$$

\z $K$ acts on functions of $x$ and $\theta$ , periodic in $\theta$, i.e.\ on the extended
Hilbert space $\mathcal{H}\otimes L_2(S,T^{-1}d\theta)$. Let now $\phi(x,\theta)$ be an
eigenfunction satisfying 

\begin{equation}
  \label{eq:link}
    K\phi=\mu\phi\ ,\ \  \phi(x,\theta+T) = \phi(x,\theta)
\end{equation}

\z then,
%
%$$ U_F\phi=e^{-i\lambda T}\phi$$
%
%\z and

$$\psi(x,t)=e^{-i\mu t}\phi(x,t)$$

\z is a solution of the Schr{\"o}dinger equation $i {\frac{\partial \psi} 
{\partial t}} = H(t) \psi$.

The existence of a real eigenvalue $\mu$ of the QEO with an 
associated $\phi(x,\theta)\in L_2({\RR}^d\otimes S)$ is thus seen to imply
the existence of a solution of the time-dependent Schr\"odinger equation
which is, in absolute value, periodic. This shows that, for
appropriate initial conditions, the particle has a nonvanishing
probability of staying in a compact domain and thus, for the case considered here, that ionization is
incomplete.  We also note that for each such $\mu$ there is actually a
whole set $\mu_n = \mu + n\omega$ of eigenvalues of $K$.  

For the specific model considered here, (\ref{eq:link}) takes the form 
\begin{equation}
  \label{eq:QEO1}
 K\phi= -\frac{\partial^2\phi(x,\theta) }{\partial
   x^2}-2(1+\eta(\theta))\delta(x)\phi -i\frac{\partial\phi }{\partial\theta}=\mu\phi
\end{equation}

\z We can now look for solutions of (\ref{eq:QEO1}) in the form

$$\phi_\mu(x,\theta)=\sum_{n\in\ZZ}y_n e^{in\omega\theta}e^{\alpha_n x}$$

\z with $\alpha_n^{\pm}=\pm\sqrt{\mu-n\omega}$. Such a
solution is in $L^2$ only if $\Re(\alpha_n x)<0$, a condition which obviously
selects different roots $\lambda_n$ depending on whether $x>0$ or $x<0$.
The requirement that $\phi_\mu$ be in $L^2(\RR)$  leads 
to a set of matching conditions which determine whether such eigenvalues 
$\mu$ can exist. It is easy to see that $\phi_\mu$ has to be
continuous at zero and satisfy the condition

$$2
\phi_\mu(0^-,\theta)-\phi_\mu(0^+,\theta)=2(1+\eta(\theta))\phi_\mu(0,\theta).
$$

\z This implies, after taking the Fourier coefficients of both sides of
the above 
equality,  the recurrence relation

\begin{equation}
  \label{eq:eqq}
  y_n(2-\alpha_n^++\alpha_n^-)=2\sum_{j\ne 0}C_jy_{n-j}
\end{equation}

\z for which a (nontrivial) solution $y_n\in l^2$ is sought. This is
effectively the same equation as (\ref{eq:homg}) below which is at the
core of our analysis.  Complete ionization thus corresponds to the
absence of a discrete spectrum of the QEO operator and conversely
stabilization implies the existence of such a discrete spectrum.  In
fact, an extension of Theorem~\ref{T2} shows that for the initial
condition $\psi_0 = u_b$, $\psi_t$ approaches such a function with $\mu
= -s_0$.  More details about Floquet theory and stability can be found
in \cite{[30],[31]}

6.  We are currently investigating extensions of our results to the case where $H_0=-\nabla^2+V_0(x)$,
$x\in\RR^d$, has a finite number of bound states and the perturbation is
of the form $\eta(t)V_1(x)$ and both $V_0$ and $V_1$ have compact
support.   Preliminary results
indicate that, with much labor, we shall be able to generalize Theorem\ref{T1},
to generic  $V_1(x)$.  The definition of genericity will, however,
depend strongly on $V_0$.

The physically important case of an external electric dipole field,
$V_1(x) = -Ex$ can be transformed into the solution of a Schr{\"o}dinger
equation of the form $H(t) = -\nabla^2 + V_0(x-g(t))$, see \cite{[2]}
This should, in principle, also be amenable to our methods but so far we have
no results for that case.

{\bf Outline of the technical strategy}.  
The method of proof relies on the properties of the Laplace
transform of $Y$, 
$y(p)=\mathcal{L}Y(p)=\int_0^{\infty}e^{-pt}Y(t)dt$.

Since the time evolution of $\psi$ is unitary,  $|\theta(t)|\le 1$.
This gives some a priori control on $Y$.  For our purposes however it is
useful to characterize directly the solution of the convolution equation
(\ref{eq:(5)}). (We restrict ourselves to $\Theta_0(k)=0$ and $I(t)=1$
there.) We show that this equation has a unique solution in suitable
norms. This solution is Laplace transformable and the Laplace transform
$y$ satisfies a linear functional equation.

 The solution of the functional
equation satisfied by the transform of $Y$ is unique in the right half
plane provided it satisfies the additional property that $y(p_0+is)$ is
square integrable in $s$ for any $p_0>0$. Any such solution $y$
transforms back (by the standard properties of the inverse Laplace
transform) into a solution of our integral equation with no faster than
exponential growth; however there is a unique locally integrable
solution of this equation, and this solution is exponentially
bounded. This must thus be our $Y$.  We can thus use the functional
equation to determine the analytic properties of $y(p)$. 

This is done using (appropriately refined versions
of) the Fredholm alternative. After some transformations, the functional
equation reduces to a linear inhomogeneous recurrence equation in $l_2$,
involving a compact operator depending parametrically on $p$, see e.g.\
(\ref{eq:rec3}).  The dependence is analytic except for a finite set of
poles and square-root branch-points on the imaginary axis and we show
that the associated homogeneous equation has no nontrivial solution. We then show
that the poles in the coefficients do not create poles of $y$, while the
branch points are inherited by $y$. The decay of $y(p)$ when
$|\Im(p)|\rightarrow\infty$, and the degree of regularity on the
imaginary axis  give us the needed information about the decay of
$Y(t)$ for large $t$.

\section{Behavior of $y(p)$ 
in the open right half plane $\mathbb{H}$}

\begin{Lemma}

\label{expEstim}

(i) Equation (\ref{eq:(5)}) has a unique solution $Y\in L^1_{loc}(\RR^+)$
and $|Y(t)|< K e^{B t}$ for some $K$, $B \in\RR$.

(ii)   The function $y(p)=\mathcal{L}Y$ exists and is analytic in 
$\mathbb{H}_{B}=\{p:\Re(p)>B\}$.

(iii)
  In $\mathbb{H}_B$, the function $y(p)$ satisfies the functional
  equation

\begin{align}\label{fnceq}
  y=  \sum_{j=-\infty}^{\infty}C_j\mathcal{T}^j\Big(h+b y\Big)
\end{align}

\z with

$$\Big(\mathcal{T} f\Big)(p)=f(p+i\omega),\ 
  h(p)=-p^{-1}\ \mbox{ and }\ \
  b(p)=-\frac{i}{p}\Big(1+\sqrt{1-ip}\Big)$$

\z 

The branch of the square root is such that for $p\in\mathbb{H}=\{p:\Re(p)>0\}$, the
real part of $\sqrt{1-ip}$ is nonnegative and the imaginary part
nonpositive.

\end{Lemma}

The straightforward proofs of this lemma are done in 
Appendix A. (Some of the results can also 
be gotten directly from standard results on the Schr\"odinger operators
and on integral equations.)

\begin{Remark}
  It is clear that  the functional equation (\ref{fnceq}) only
  links  points on the  one dimensional lattice $\{p+i\ZZ\omega\}$. It
  is convenient to take $p_0$ such that $p=p_0+i n\omega$ with
  $\Re(p_0)=\Re(p)$ and 
\begin{equation}
  \label{eq:norm}
  \Im(p_0)\in [0,\omega)
\end{equation}

\end{Remark}
\z The functions ${y}, {h},\, {b}$ in (\ref{fnceq}) will now depend
parametrically on $p_0$.  We set ${y}=\{y_j\}_{j\in\ZZ}$,
${h}=\{h_j\}_{j\in\ZZ}$, ${b}= \{b_j\}_{j\in\ZZ}$ with $y_n=y(p_0+i n\omega)=y(p)$
(and similarly for $h(p)$ and $b(p)$).  It is convenient to define the
operator $(\hat{H} y)_n=b_{n}y_n$. Let $(\mathcal{T} y)_n=y_{n+1}$ be
the right shift on $l_2(\ZZ)$ (which we denote for simplicity by $l_2$)
and rewrite (\ref{fnceq}) as

\begin{equation}
  \label{eq:rec3} {y}=
\sum_{j=-\infty}^{\infty}C_j\mathcal{T}^j{h}+\sum_{j=-\infty}^{\infty}C_j\mathcal{T}^j\hat{H}{y}\equiv
{f}+\mathcal{J}{y}
\end{equation}

\begin{Proposition}
  \label{L2}
  For $\Re(p_0)> 0$ there exists a unique solution of (\ref{eq:rec3}) in
  $l_2$. This solution is analytic in $p_0, \Re(p_0)>0$.  Thus
  $y(p)$ is analytic in $p\in\mathbb{H}$ and inverse Laplace
  transformable there with $\mathcal{L}^{-1}(y)=Y$.

\end{Proposition}

 {\it Proof}
The proof uses the Fredholm alternative. We first prove the following 
results. 

\begin{Lemma}\label{Pcompact}
  The operator $\mathcal{J} $ is
  compact on $l_2$ if $p_0\ne 0$. 
\end{Lemma}
\renewcommand{\labelenumi}{(\Alph{enumi})}

\begin{proof} 
  The proof uses standard compact operator results, see e.g.
  \cite{Reed-Simon}.  \emph{First note that the operator
    $\mathcal{T}^j\hat{H}$ is compact.}  This is straightforward:
  since $b_{j}\rightarrow 0$ as $j\rightarrow\infty$, it follows that
  $\hat{H}$ is the norm limit as $N\rightarrow\infty$ of the finite
  rank operators defined by $(\hat{H}_N y)_j =b_{j}y_j$ for $|j|\le N$
  and $(\hat{H}_N y)_j=0$ otherwise, and thus is compact.  Since
  $\|\mathcal{T}\|=1$, $\mathcal{T}^j\hat{H}$ is compact too.  As the
  series $\sum_{j=-\infty}^{\infty}C_j\mathcal{T}^j\hat{H}$ converges
  strongly, $\mathcal{J} $ is compact.

\end{proof}

{\bf{Remarks}}

{\bf{1.}} Note that ${f}\in l_2$ if $p_0\ne 0$ (a
straightforward consequence of the fact that $\mathbf{C}$ and $h$ in
(\ref{eq:rec3}) are in $l_2$).

{\bf{2.}} The operator $\mathcal{J}$ is analytic in $p_0$, except for
$p_0=0$, where the coefficients have poles, and for an additional value
on the imaginary axis (possibly also $0$), where the coefficients have square
root branch points.

\begin{Remark}\label{P7}

Setting, for $p_0\ne 0$,  
 \begin{equation}
    \label{eq:defyz}
    y_l=(\sqrt{1-i(p_0+i l\omega)}-1)z_l
  \end{equation}
the homogeneous equation
\begin{equation}
  \label{eq:eqy1}
y=\mathcal{J}y  
\end{equation}

\z clearly has a (nontrivial) $l_2$ solution $y$  only if

\begin{equation}
  \label{eq:homg}
 \Big(\sqrt{1-ip_0+l\omega}-1\Big)z_l=- \sum_{k=1}^{\infty} \Big(C_k z_{l+k}+\overline{C}_k z_{l-k} \Big)
 \end{equation}
\z has a (nontrivial) $l_2$ solution $z$ with

\begin{equation}
  \label{eq:eqzz}
  \left\{\Big(\sqrt{1-ip_0+j\omega}-1\Big)z_j\right\}_{j\in\ZZ}\in l_2
\end{equation}

\end{Remark}

\begin{Lemma}\label{P1} 

 For any $\eta$ under assumptions (a) to (c), if $p_0\in {\mathbb{H}}$
  there is no nonzero $l_2$ solution of (\ref{eq:homg}) such that
(\ref{eq:eqzz}) holds.

\end{Lemma}

\begin{proof}
  
  To get a contradiction, assume $z\in l_2$, $z\not\equiv 0$,
  satisfying (\ref{eq:eqzz}), is a solution of (\ref{eq:homg}).
  Multiplying (\ref{eq:homg}) by $\overline{z_l}$, and summing with
  respect to $l$ from $-\infty$ to $+\infty$ we get

\begin{multline}
  \label{eq:sum}
\sum_{l=-\infty}^{\infty}  \Big(\sqrt{1-ip_0+l\omega}-1\Big)|z|^2_l
 \\=- \sum_{l=-\infty}^{\infty}\sum_{k=1}^{\infty} \Big(C_k
z_{l+k}\overline{z_l}+\overline{C}_k z_{l-k}\overline{z_l} \Big)=
-\sum_{l=-\infty}^{\infty}\sum_{k=1}^{\infty}\Big(C_k
z_{l}\overline{z_{l-k}}+\overline{C}_k z_{l-k}\overline{z_l}\Big)\\=
-\sum_{l=-\infty}^{\infty}\sum_{k=1}^{\infty}2\Re\Big(C_k
z_{l}\overline{z_{l-k}}\Big)
\end{multline}

\z  If $p_0\in {\mathbb{H}}$ the imaginary part of
$\sqrt{1-ip_0+l\omega}$ is negative (see (\ref{intres}) and the
discussion following it) and thus, if some $z_l$  is nonzero
then the left side of (\ref{eq:sum}) has strictly negative imaginary
part, which is impossible since the right side is real.  
\end{proof}

{\it Proof of Proposition~\ref{L2}} The existence of the analytic
solution follows now immediately from the analytic Fredholm alternative
and the analyticity of the coefficients, for $p_0\in\mathbb{H}$. The
fact that $\{y_n\} \in l_2$ together with the stated analyticity imply
that the function $\mathcal{L}^{-1}y(p)$ exists and satisfies the
  integral equation of $Y$, and thus coincides with $Y$.

\section{Behavior of $y(p)$ in the neighborhood of $\Im(p)=0$  in the generic case}\label{S5}

{\bf Discussion of methods}. We start again from relation
(\ref{eq:rec3}).  This has the form 

\begin{equation}
  \label{eq:eqyv}
  y_n=i\sum_j \frac{C_j}{-ip_0+(n+j)\omega}-\sum_jC_jq_{n+j}y_{n+j}\ \ ,\ \  \  C_0=0
\end{equation}

\z where

\begin{equation}
  \label{eq:qk}
q_n=\frac{
[1+\sqrt{1-ip_0+n\omega}]}{-ip_0+n\omega}  
\end{equation}

\z As the imaginary axis $\Re(p_0)=0$ is approached, two types of
potential
singularities in the coefficients need attention: the poles in the coefficients due to 
the presence of $p^{-1}$, and the square root singularities. It will
turn out that by cancelation effects, the poles play  no
role, generically. The square root singularities will be manifested in the solution
$y$. 
The study of these questions requires further regularization of the
functional
equation  (\ref{eq:eqyv}).

It is convenient to separate out the terms in (\ref{eq:eqyv}) which are
singular at $p_0=0$. Using (from now on) the notation $s_0=-ip_0$ we have

\begin{multline}
  \label{eq:yv0}
  y_n=i\frac{C_{-n}}{s_0}-\frac{C_{-n}(1+\sqrt{1+s_0})}{s_0}y_0
+i\sum_{j\ne -n}\frac{C_j}{s_0+(n+j)\omega}\\
-\sum_{j\ne
  -n}C_jq_{n+j}y_{n+j}, \quad n \ne 0
\\
 y_0=i\sum_{j\ne 0} \frac{C_j}{s_0+j\omega}-\sum_{j\ne 0}C_jq_jy_j\qquad\qquad\qquad\qquad
\end{multline}

We break up the proof into two parts, the non-resonant and resonant case.  We
start with the former,

\subsection{ The Non-Resonant Case, $\omega^{-1}\not \in \NN $}

\begin{Proposition}\label{yanat0}
  If condition (g) is satisfied, and $\omega^{-1} \not\in \NN$, then the solution $y$ of
  (\ref{eq:yv0}) is analytic in a small neighborhood of $s_0=0$.
\end{Proposition}

For the proof we write $y_0=i/2+s_0 u_0$,
%Then
%
%$$\frac{i}{s_0}-\frac{1+\sqrt{1+s_0}}{s_0}y_0=
%\frac{i(1-\sqrt{1+s_0})}{2s_0}-(1+\sqrt{1+s_0})u_0$$
%
%\z so that
%                
%\begin{align}
%  \label{eq:14}
%  y_n&=f_n-C_{-n}(1+\sqrt{1+s_0})u_0-\sum_{k\ne
%    0}q_kC_{k-n} y_k \ \ (n\ne 0)
%\end{align}
%
%\z and
%
%\begin{equation}
% s_0\,u_0=f_0
%-\sum_{j\ne 0}C_j q_j y_j
%%%%%\end{equation} 
and for $n\ne 0$ we make the substitution $y_n=v_n+d_n u_0$, where we
will choose $d_n$ according to (\ref{eq:dn}) in order to eliminate
$u_0$ from all equations with $n\ne 0$.

%We have
%
%\begin{align}\label{sdv}
%y_n =   v_n+d_n u_0&=f_n-C_{-n}(1+\sqrt{1+s_0})u_0-\sum_{k\ne
%    0}C_{k-n}q_kv_k-\sum_{k\ne 0}C_{k-n}q_kd_ku_0\ \ ,\ \  n
%\ne 0 \\
%y_G = \frac{1}{2} + s_0 u_0&=f_0-\sum_{j\ne 0}C_jq_jv_j-\sum_{j\ne 0}C_jq_j d_j u_0, 
%\end{align}

\begin{Lemma}\label{P33}
(i)  For $s_0\in\RR$ there exists a unique solution $d\in l_2(\ZZ\setminus\{0\})$ 
of the system 

\begin{equation}
  \label{eq:dn}
  d_n=-C_{-n}(1+\sqrt{1+s_0})-\sum_{k\ne 0}C_{k-n}q_k d_k,\ \ n\ne 0
\end{equation}

\z This solution is analytic at $s_0=0$.

(ii) With this choice of $d$, the system (\ref{eq:yv0}) becomes

\begin{align}\label{eqvn}
  v_n&=f_n-\sum_{k\ne 0}C_{k-n}q_kv_k \nonumber \\
\left(s_0+\sum_{j\ne 0}C_jq_jd_j\right)u_0&=f_0-\sum_{j\ne 0}C_j q_j v_j
\end{align}

\z where

\begin{align}\label{28}
f_0=-\frac{i}{2}+i\sum_{j\ne 0}\frac{C_j}{s_0+j\omega},\   f_n=i C_{-n}\frac{1-\sqrt{1+s_0}}{2s_0}+i\sum_{k\ne
    0}\frac{C_{k-n}}{s_0+k\omega}
\end{align}

(iii) For small $s_0$ we have $ \sum_{j\ne 0}C_jq_jd_j\ne 0$, and the
system (\ref{eqvn}) has a unique solution with $v\in l_2(\ZZ\setminus\{0\})$,
and $v_n$, $u_0$ are analytic at $s_0=0$ .
\end{Lemma}

\begin{proof}
  (i) Equation (\ref{eq:dn}) is of the form $(\mathbb{I}-\mathcal{J}')d=c'$
in $l_2(\ZZ\setminus\{0\})$, 
where $c'_n=-(1+\sqrt{1+s_0})\overline{C}_n$ and

$$(\mathcal{J}'d)_n=-\sum_{k\ne 0}C_{k-n}q_k d_k,\ \ \ (n\ne 0)$$

We show first that $\mbox{Ker} (\mathbb{I}-\mathcal{J}')=\{0\}$. Indeed, let $d=Ad$
and set $D_k =q_k d_k$. Then  we see that

\begin{equation}
  \label{eq:D}
  {q_n}^{-1}D_n+\sum_{k\ne 0}C_{k-n}D_k=0
\end{equation}

\z and, by multiplying with $\overline{D}_n$ and summing over $n$ we get

\begin{equation}
  \label{eq:D1}
  \sum_{n\ne 0}q_n^{-1}|D_n|^2+\sum_{n,k\ne 0}C_{k-n}D_k \overline{D_n}=0
\end{equation}

\z Note that, because $C_{-n}=\overline{C}_n$, the following quantity is real:

\begin{equation}
  \label{eq:cc}
  \overline{\sum_{n,k\ne 0}C_{k-n}D_k\overline{D}_n}=
\sum_{n,k\ne 0}C_{n-k}\overline{D_k} D_n=\sum_{n,k\ne
  0}C_{k-n}D_k \overline{D}_n 
\end{equation}

\z implying that 

$$\sum_{n\ne 0}q_n^{-1}|D_n|^2\in\RR$$

\z with (cf. (\ref{eq:qk}))

$$q_n^{-1}=-1+\sqrt{1+s_0+n\omega}$$

\z Let $N_0=-(1+s_0)\omega^{-1}\in\RR$. Obviously $q_n^{-1}\in\RR$ for $n\ge
N_0$ while for $n<N_0$ we have, by Lemma~\ref{expEstim} (iii)

$$\Im(q_n^{-1})<0$$

\z Thus it is necessary that $D_n=0$ for all $n<N_0$. 

Assume $D\ne 0$. Let $N\in\NN$ be such that $D_n=0$ for all $n<N$ and
$D_N\ne 0$ (thus $N_0\leq N$). Then from (\ref{eq:D})
$$\sum_{k\ge N;k\ne 0}C_{k-n}D_k=0\ \ \ \mbox{ for any }n<N$$ 
or, setting $k=N-1+j$,
\begin{equation}\label{xxDj}
\sum_{j\geq 1,j\ne 1-N}C_{j+n}D_{N-1+j}=0\ \ \ \ {\mbox{for}}\ \
n\geq 0 
\end{equation}

It is here that we use the genericity condition on $\bf C$. In fact we
will show that (\ref{xxDj}) implies $D=0$ if condition (g) is
satisfied.  To see this define $\tilde{D}\in l_2(\NN)$ as
$\tilde{D}_j=D_{N-1+j}$ if ${j\geq 1,j\ne 1-N}$ and, if $1-N\geq 1$,
$\tilde{D}_{1-N}=0$.  Then by (\ref{xxDj}) $\tilde D$ is orthogonal in
$l_2(\NN)$ to all $T^nC$, $n\geq 0$.  \z By the genericity condition
(g) then $<\tilde{D},e_1>=D_N=0$, which is a contradiction. Thus
$D=0$.

Since $\mathcal{J}'$ is analytic in $s_0$ for small enough $s_0$, and compact by
the same simple arguments as in Lemma~\ref{Pcompact} it follows
that $(\mathbb{I}-\mathcal{J}')^{-1}$ exists and is analytic in $s_0$ at $s_0=0$.

(ii) This part is an immediate calculation.

(iii) Note first that $f\in l_2(\ZZ\setminus \{0\})$, because

$$\|f\|\le \left|\frac{1-\sqrt{1+s_0}}{2s_0}\right|\|c\|+
\left(\sum_{n\ne 0}\left|\sum_{k\ne
      0}\frac{C_{k-n}}{s_0+k\omega}\right|^2
\right)^{1/2}\le \|c\|\sum_{k\ne 0}\frac{1}{|s_0+k\omega|^2}<\infty
$$

\z Also, $f$ being an integral with respect to a discrete measure of an
$l_1$ analytic function, depends analytically on small $s_0$.

The rest of the proof of (iii)  closely follows that of part (i),
using the following result.

\begin{Lemma}

For $s_0=0$ we have  $\displaystyle \sum_{j\ne 0}C_jq_jd_j\ne 0$.

\end{Lemma}

\begin{proof}
  Assume the contrary was true. At $s_0=0$, with
$D_n^0={D_n}|_{s_0=0}$ and $q_n^0=q_n|_{s_0=0}$, relation
(\ref{eq:D}), using (\ref{eq:dn}), gives

  \begin{align}\label{orel}
    0&= \frac{D^0_n}{q^0_n}=-\sum_{k\ne 0}C_{k-n}D^0_k-2C_{-n} \ \ 
    (n\ne 0)
  \end{align}
\z Multiplying with $\overline{D^0_n}$ and summing over $n\ne 0$ 
we would get 
\begin{equation}
  \label{eq:contr}
  \sum_{n\ne 0}(-1+\sqrt{1+n\omega})|D^0_n|^2
=-\sum_{k,n\ne 0}C_{k-n}D^0_k\overline{D^0_n}-\sum_{n\ne 0}2C_{-n}\overline{D^0_n}
\end{equation}

\z and since we assumed $\sum_n C_nD^0_n=0$ then, as in the proof of
Lemma~\ref{P33} (i), it follows that $D^0_n=0$ for all $n<
N_0=-\omega^{-1}$. This gives, using (\ref{orel}), that

\begin{equation}
  \label{eq:Dc}
  \sum_{k\ge N_0;k\ne 0}C_{k-n}D^0_k+2C_{-n}=0
\end{equation}

Denote by ${D}^1\in l_2$ the sequence  ${D}^1_k=D^0_k$ if $k\ne 0$ and
${D}^1_0=2$. As in the proof of Lemma~\ref{P33} (i), using the
genericity condition (g), we get ${D}^1=0$, an obvious contradiction.
\end{proof}
\end{proof}

This concludes the proof of Proposition~\ref{yanat0}: for generic $\eta$
the solution $y$ of (\ref{eq:rec3}) has, for $\omega^{-1}\notin\NN$,
analytic components $y_n$ when $p=0$.

\z {\bf Square root singularities}. We now study the behavior at the
square root singularities of the coefficients of the equation of $y$.

Let  $k_0$  be the unique integer such that for some $s_r\in [0,\omega)$
we have  $1+s_r+k_0\omega=0$ (then $s_r$ is a branch point in the
coefficient $q$).

The following Proposition describes the analytic structure of $y(p)$ near
the imaginary axis.

\begin{Proposition}
We have the decomposition
$y_n=u_n+(\sqrt{s_0-s_r})v_n$ where $u_n$ and $v_n$ are analytic in 
$s_0$ in a {\em complex} neighborhood of the segment $[0,\omega)$.

\end{Proposition}

\begin{proof}
  The substitution $y_n=u_n+(\sqrt{s_0-s_r})v_n$, and
$$U_k=q_ku_k;\
V_k=q_kv_k\ \ (k\ne k_0)\
\mbox{and}\ 
U_{k_0}=\frac{u_{k_0}}{s_0+k_0\omega};\
\
V_{k_0}=\frac{v_{k_0}}{s_0+k_0\omega}$$

\z leads to the following
  system of equations for $U_n$ and $V_n$

  \begin{align}\label{s6}
    q_n^{-1}U_n&=ri\sum_{k}\frac{C_{k-n}}{s_0+k\omega}-\sum_{k}C_{k-n}U_k
-C_{k_0-n}(s_0-s_r)V_{k_0}\ \ (n\ne k_0)\nonumber \\
 q_n^{-1}V_n&=-\sum_{k}C_{k-n}V_k
-C_{k_0-n}(s_0-s_r)V_{k_0}-C_{k_0-n}U_{k_0}\ \ (n\ne k_0)\nonumber \\
(s_0+k_0\omega)U_{k_0}&=i\sum_k\frac{C_{k-k_0}}{s_0+k\omega}\nonumber \\
(s_0+k_0\omega)V_{k_0}&=-\sum_{k}C_{k-k_0}V_k
  \end{align}

\z We now let $Q_{k_0}=s_0+k_0\omega$ and, for $n\ne k_0$,
$Q_n=q_n^{-1}=-1+\sqrt{1+s_0+k\omega}$.
 We use again the Fredholm alternative and, as in the previous proofs,
 we need only to show
the absence of a solution of the homogeneous equation {\em at}
$s_0=s_r$. We thus multiply the homogeneous equations associated to
(\ref{s6}) in the following manner: the equation for $U_j$ by
$\overline{U_j}$ and the equation for $V_j$ by $\overline{V_j}$, then sum
over all $j$.  As in the previous proofs, from the reality of the
r.h.s. and then from the genericity condition (g)  $U\equiv 0$.  Then,
similarly,  $V\equiv
0$. The rest is immediate.
\end{proof}

\subsection{The resonant case: $\omega^{-1}=M\in\NN$}

In this case when $s_0=0$ there are poles in the coefficients of
(\ref{eq:eqyv}) when $n+j=0$
and branch points when $n+j=-M$. The proof is a combination of the two
regularization techniques used in the previous case.

\begin{Proposition}\label{resonant}
 We can set $y(s_0)=A(s_0)+B(s_0)\sqrt{s_0}$ with $A$ and $B$ analytic in a
  complex
neighborhood of the segment $[0,\omega)$.
\end{Proposition}

\begin{proof}
Special care is only needed near $s_0=0$.  The system
(\ref{eq:dn})--(\ref{28}) now reads

  \begin{align}
    d_n&=-C_{-n}(1+\sqrt{1+s_0})-\sum_{k\notin\{0,-M\}}C_{k-n}q_k d_k-
C_{-M-n}\frac{1+\sqrt{s_0}}{s_0-1}d_{-M}
\nonumber\\v_n&=f_n-\sum_{k\notin\{0,-M\}}C_{k-n}q_k v_k-C_{-M-n}\frac{1+\sqrt{s_0}}{s_0-1}v_{-M}
  \end{align}

\z We take $d_n=\alpha_n+\sqrt{s_0}\beta_n$ and
$v_n=\gamma_n+\sqrt{s_0}\delta_n$. The system becomes

\begin{align}
  \alpha_n&=-C_{-n}(1+\sqrt{1+s_0})-\sum_{k\notin\{0,-M\}}C_{k-n}q_k
  \alpha_k
-C_{-M-n}\frac{1}{s_0-1}(\alpha_{-M}+s_0\beta_{-M})\nonumber \\
\beta_n&=-\sum_{k\notin\{0,-M\}}C_{k-n}q_k
  \beta_k-C_{-M-n}\frac{1}{s_0-1}(\alpha_{-M}+\beta_{-M})\label{s1}\\
\gamma_n&=f_n-\sum_{k\notin\{0,-M\}}C_{k-n}q_k\gamma_k-C_{-M-n}\frac{1}{s_0-1}
(\gamma_{-M}+s_0\delta_{-M})\nonumber\\
\delta_n&=-\sum_{k\notin\{0,-M\}}C_{k-n}q_k
  \delta_k-C_{-M-n}\frac{1}{s_0-1}(\delta_{-M}+\gamma_{-M})\label{s*2}
\end{align}

\z The system (\ref{s1}) is of the form 

$$\begin{pmatrix} \alpha\\ \beta \end{pmatrix} =S(s_0)\begin{pmatrix}
  \alpha\\ \beta \end{pmatrix}+\begin{pmatrix} F_1 \\ F_2 \end{pmatrix}$$

\z where $\alpha, \beta, F_1, F_2$ are in $l_2$. We prove that
the homogeneous equation has no nontrivial solutions:

\begin{Lemma}
  \label{triv}
$(\mathbb{I}-S(0))\begin{pmatrix} \alpha\\ \beta \end{pmatrix}=0$
implies $\begin{pmatrix} \alpha\\ \beta \end{pmatrix}=0$. 
\end{Lemma}

\begin{proof}
  Let $Q_n=q_n$,  $A_n=q_n\alpha_n$, $B_n=q_n\beta_n$ for $n\ne 0, -M$
  and $Q_{-M}=-1$, $A_{-M}=-\alpha_{-M}$ and $B_{-M}=-\beta_{-M}$. 
The system (\ref{s1}) becomes

\begin{align}
  Q_n^{-1}A_n&=-\sum_{k\ne 0}C_{k-n}A_k\nonumber\\
 Q_n^{-1}B_n&=-\sum_{k\ne 0}C_{k-n}B_k -C_{-M-n}A_{-M}
\end{align}

\z As in the proofs in Case I, multiplying the first equation by
$\overline{A_n}$, summing over $n$ we first get from the reality
of the r.h.s. that $A_{n}=0$ for $n<-M$ and then by the condition (g)
we get that $A\equiv 0$. The conclusion $B\equiv 0$ now follows in the
same way.

\end{proof}

\z {\em End of proof of Proposition~\ref{resonant}}. The operator $S$ is
compact on $l_2\oplus l_2$ and $S$ and $(F_1,F_2)$ are
analytic in a complex neighborhood of $0$. We saw in
Lemma~\ref{triv} that the kernel of $\mathbb{I}-S(0)$ is trivial
and by the analytic Fredholm alternative it follows that
$(\mathbb{I}-S(0))^{-1}$ exists and is analytic in a small neighborhood of
$s_0=0$. Hence $(\alpha,\beta)$ are analytic. Similarly, $\gamma,\delta$
are analytic in the same region.
\end{proof}

\subsection{Proof of Theorem \ref{T1}}
Combining the above results we have the following conclusion:
\begin{Proposition}
 If condition (g) is fulfilled, then $y(p)$ is analytic in a neighborhood of
 $i\RR\setminus\{is_r+i\omega\ZZ\}$. For any
 $j\in\ZZ$, in a neighborhood
of $p=is_r+ij\omega$ ($s_r\in\RR$) $y$ has the form
$y(p)=A_j(p)+B_j(p)\sqrt{-ip-s_r-ij\omega}$ where $A_j$ and $B_j$ are
analytic. In particular, $y$ is Lipschitz continuous of exponent 
$1/2$ in the closed right half plane. Thus $Y(t)=O(t^{-3/2})$
for large $t$.
\end{Proposition}

\begin{proof}
  All but the last claim has already been shown. The last statement
is a standard Tauberian theorem (note that $\mathcal{L}^{-1}$
is the Fourier transform along the imaginary line).
\end{proof}
\begin{Proposition}
  We have $\theta(t)\rightarrow 0$ as $t\rightarrow\infty$.
\end{Proposition}

\begin{proof} We can write (\ref{eq:(5)}) (with $I(t)=1$) as

  \begin{equation}
    \label{eq:neweq}
    Y=\eta(\theta+M*Y)
  \end{equation}
  It is easy to check, in view of the fact that $M$ and $Y$ are
  $O(t^{-3/2})$, that $M*Y\rightarrow 0$.  Furthermore $1+2i\int_0^t
  Y(s)ds$ is convergent as $t\rightarrow\infty$. Thus
  $\theta(t)\rightarrow const$ as $t\rightarrow\infty$.  Since now the
  l.h.s.  of (\ref{eq:(5)}) converges to zero and $\eta$ does not, the
  equality (\ref{eq:neweq}) is only consistent if $\theta(t) \to 0$.

\end{proof}
This completes the proof of Theorem \ref{T1} for the case $\psi_0 = u_b
= e^{-|x|}$.

The general case follows by noting that the inhomogeneous term does not
affect the main argument, using the Fredholm alternative.  Hence we will
still have $|\theta(t)| \to 0$ but the rate of decay may be different.

\section{A nongeneric example} 

Let $\eta$ be given by (\ref{eq:counterex}), for which
\begin{equation}\label{coefCn}
C_n=-r\lambda^n\ \ {\mbox{for}}\ n\geq 1\ \ \ \ \ ,\ \ \ C_n=C_{-n}
\end{equation}

As in \S~5 set $-ip_0=s_0$ and let $q_n$ be given by
(\ref{eq:qk}). Denote 
\begin{equation}\label{an}
a_n=a_n(s_0)=\frac{1}{r}\frac{1}{q_n}=\frac{1}{r}\left(\sqrt{1+s_0+n\omega}-1\right)
\end{equation}

For $r\in(0,1)$, $\omega>1$, $\omega^{-1}\not\in\NN$ such that
$(1-r)^2<\omega -1$, let $s_r$ and $s_p$ be the unique numbers in $(0,\omega)$ so
that $1+s_r\in\omega\ZZ$ and $1+a_{-1}(s_p)=0$. We choose $r,\omega$ such that $s_r\ne s_p$.

\subsection{The homogeneous equation}

\begin{Lemma}\label{ancompo}
  
  Let $s_{0,0}$ be a point in $(0,s_r)\cup(s_r,\omega)$. Consider $s_0$ in a small
  enough neighborhood of $s_{0,0}$. The linear operator
  $\mathcal{J}=\mathcal{J}(s_0)$ of (\ref{eq:rec3}) depends
  analytically on $s_0$, and is compact on $l_2$. For $s_0\ne s_p$,
  $(I-\mathcal{J}(s_0))^{-1}$ exists and is analytic.
\end{Lemma}

\begin{Lemma}\label{homogsol}

Denote for short $\mathcal{J}_0=\mathcal{J}(s_r)$. 
  
There exists a value $\lambda=\lambda_s\in(0,1)$ such that
\begin{equation}
\dim {\mbox{Ker}}\left(I-\mathcal{J}_0\right)=1
\end{equation}

\end{Lemma}

Denote by $A$ the diagonal (unbounded) operator $(Az)_n=a_nz_n$ in
$l_2$; $A^{-1}$ is bounded.

\begin{Lemma}\label{keradj}

For $\lambda=\lambda_s$ as in Lemma \ref{homogsol} we have
\begin{equation}
  {\mbox{Ker}}\left(I-\mathcal{J}_0\right)=
A\left[ {\mbox{Ker}}\left(I-\mathcal{J}_0^*\right)\right]
\end{equation}

\end{Lemma}

\subsection{Proof of Lemma \ref{ancompo}}

The operator $\mathcal{J}$ is compact by Lemma \ref{Pcompact}. To show
that $I-\mathcal{J}$ is invertible we prove this for any points
$s_0\in(0,\omega)$, $s_0\ne s_p,s_r$; by the analytic Fredholm theorem
it will follow that $I-\mathcal{J}$ is invertible in a small enough
neighborhood of any such point, thus proving the Lemma.

Let $s_0\in(0,\omega)$, $s_0\ne s_p,s_r$. As in Remark \ref{P7} in
section \ref{S5}, the
substitution $y_n=a_nz_n$ ($n\in\ZZ$) transforms the homogeneous
equation (\ref{eq:eqy1}) to
\begin{equation}\label{recuzngen}
a_n z_n=\sum_{j=1}^{\infty}\lambda^j\left(z_{n+j}+z_{n-j}\right)\ \ \ ,\ n\in\ZZ
\end{equation}
\z Note that $\Im a_n<0$ for $n<-1$ for $s_0\in[\omega-1,\omega)$ and
$\Im a_n<0$ for $n<0$ for $s_0\in(0,\omega-1)$. We will discuss only
the first case, $s_0\geq \omega-1$, since the second one is completely
analogous. 

As in the proof of Lemma~\ref{P1}, it follows that
\begin{equation}\label{whichzn0}
z_n=0\ \ \ {\mbox{\ for}}\ \  n<-1
\end{equation}
Then equations (\ref{recuzngen}) for $n<-1$ become
\begin{equation}\label{eq:syst5}
\sum_{k=1}^\infty\lambda^kz_{k-2}=0
\end{equation}
\z For $n={-1}$ (\ref{recuzngen}) gives 
\begin{equation}\label{zm1is0}
(a_{-1}+1)z_{-1}=0
\end{equation}
and for $n\geq 0$, using (\ref{eq:syst5}), we get

\begin{equation}\label{eq:syst4}
\displaystyle (1+a_n) z_n=\sum_{j=1}^{n+1}(\lambda^j-\lambda^{-j})
z_{n-j}\ \ \ ,\ n\geq -1
\end{equation}

Since $s_0\ne s_p$, (\ref{zm1is0}) gives $z_{-1}=0$, and it follows by
induction, from (\ref{eq:syst4}), that $z_n=0$ for all $n$. By the
Fredholm alternative theorem then $I-\mathcal{J}(s_0)$ is invertible.

\subsection{Proof of Lemma \ref{homogsol}}

In what follows $s_0=s_r$.

\subsubsection{An auxiliary lemma}
We show that if $z\in l_2$ then  equation (\ref{eq:syst5}) is redundant.

\begin{Lemma}
  If $z$ is an $l_2$ solution of (\ref{eq:syst4}) with $z_n=0$ for
  $n<-1$ then $z$ satisfies (\ref{eq:syst5}).

\end{Lemma}

\begin{proof}

Let $z\in l_2$ be a solution of (\ref{eq:syst4}). Then 
\begin{equation}\label{49t}
Z^{[n+1]}\equiv\sum_{k=1}^n\lambda^kz_{k-2}
\end{equation}

\z is the truncation of a convergent series, since there is a
constant $M$ with $|z_n|<M$ for all $n$. Note that
$$(1+a_n)z_n=\sum_{j=1}^{n+1}\lambda^jz_{n-j}-\lambda^{-n-2}Z^{[n+1]}$$
hence
$$Z^{[n+1]}=\lambda^{n+2}\sum_{j=1}^{n+1}\lambda^jz_{n-j}-\lambda^{n+2}(1+a_n)z_n$$
so that
\begin{equation}\label{zn0}
\left| Z^{[n+1]}\right|  \leq
\lambda^{n+2}\frac{M\lambda}{1-\lambda}+\lambda^{n+2}M\left( 1+\,
  {\mbox{const}}\, \sqrt{n}\right)\longrightarrow 0\ \ \ {\mbox{as}}\
n\rightarrow\infty
\end{equation} 

\z Since (\ref{49t}) are truncations of the series in the LHS of
(\ref{eq:syst5}), then (\ref{zn0}) implies (\ref{eq:syst5}).

\end{proof}

\subsubsection{Behavior of the general solution of (\ref{eq:syst4})}

A direct calculation shows that the sequence $z_n$ satisfying the
infinite order recurrence (\ref{eq:syst4}) and the initial condition
$z_{-1}=1$ satisfies, in fact, the three step recurrence

\begin{multline}\label{eq:Zs}
(1+a_{n+1})
z_{n+1}+(1+a_{n-1})z_{n-1}=[\lambda(1+a_n)+\lambda+a_n\lambda^{-1}]z_{n}\
\ (n\geq 0) 
\end{multline}
with the initial condition
\begin{equation}\label{eq:Zsini}
z_{-1}=1,\ \ \ \ z_0=\frac{\lambda-\lambda^{-1}}{1+a_0}.
\end{equation}

Denote
\begin{equation}
  \label{eq:zV}
  z_n=\frac{\lambda-\lambda^{-1}}{1+a_n}V_{n-1}
\end{equation}
then (\ref{eq:Zs}) becomes
\begin{equation}
  \label{eq:V}
  V_{n}+V_{n-2}=\left[\lambda+\frac{\lambda^2+a_n}{\lambda(1+a_n)}\right]
V_{n-1} \ \ ,\ n\geq 1
\end{equation}

We are looking for $l_2$ solutions. Standard WKB (see e.g. \cite{[33]})
would imply there are solutions of the discrete equation (\ref{eq:V})
behaving like $\lambda^{-n}e^{-\sqrt{n/\omega}}$ and like
$\lambda^{n}e^{\sqrt{n/\omega}}$. We will prove this in our context and
find special values of $\lambda$ for which the solution decaying for
large $n$ satisfies the initial condition. We will show that this
solution is obtained by taking
\begin{equation}\label{VrV}
V_{n-2}=g_{n-1}V_{n-1}
\end{equation}
in (\ref{eq:V}) and iterating:

\begin{equation}
  \label{eq:contf}
g_{n-1}=G_n-\frac{1}{g_n}\mbox{ with }G_n=\lambda+\frac{\lambda^2+a_n}{\lambda(1+a_n)}  
\end{equation}

\z i.e., $g_{0}$ is given by the continued fraction:

$$g_{0}=G_{1}-\frac{\displaystyle 1}{\displaystyle G_{2}-\frac{\displaystyle 1}{\displaystyle G_3}\ldots}$$

\z which needs to match the initial condition (see (\ref{eq:Zsini}): 
\begin{equation}\label{inicd}
 g_{0}=g_0(\lambda)=\frac{1}{\lambda+\lambda^{-1}
+(1+a_0)^{-1}(\lambda-\lambda^{-1})}
\end{equation}

\begin{Lemma}\label{contf}
  (i) Let $\lambda\in (0,1)$. The recurrence (\ref{eq:contf}) has a solution such that
  $g_n\rightarrow\lambda^{-1}$ as $n\rightarrow\infty$.

 (ii) $g_{0}$ is meromorphic in $\lambda$ on $[0,1)$ and has poles.
  
  (iii) There exists $\lambda_s\in(0,1)$ such that $g_{0}(\lambda_s)$
  satisfies (\ref{inicd}).
  
  (iv) Let $\lambda=\lambda_s$. To the solution of (i) there
  corresponds a solution $V^{[s]}$ of the recurrence (\ref{eq:V}) such
  that $V_n^{[s]}\sim \lambda_s^{n+o(n)}$ as $n\rightarrow\infty$.
  The corresponding solution $z^{[s]}$ of (\ref{eq:syst4}) satisfies
  $z_n\rightarrow 0$ as $n\rightarrow\infty$.
  
  (v) Let $\lambda=\lambda_s$. There exists a solution of (\ref{eq:V})
  with the asymptotic behavior $V_n^{[l]}\sim \lambda_s^{- n+o(n)}$.
  
  Thus, for $\lambda=\lambda_s$, there exists a unique (up to a
  multiplicative constant) ``small'' solution of (\ref{eq:V}), with
  the behavior $V_n^{[s]}\sim \lambda_s^{n+o(n)}$ for large $n$, while
  the general solution behaves like $V_n\sim \lambda_s^{-n+o(n)}$. As
  a consequence, a similar statement holds for the recurrence
  (\ref{eq:Zs}).

\end{Lemma}

{\em{Remark.}} The proof of (iii) can be refined to show that, in fact,
there is a countable set of points $\lambda_s$ for which $g_0$
satisfies the initial condition, and that these values accumulate to
$1$.

\begin{proof}

(i) With the substitution
\begin{equation}\label{dclr}
g_n=G_{n+1}-\lambda+\delta_n
\end{equation}
the recurrence (\ref{eq:contf}) becomes
\begin{equation}\label{recdel}
\delta_n=\lambda-\frac{1}{G_{n+2}-\lambda+\delta_{n+1}}\equiv
\left(\mathcal{S}\delta\right)_n\ \ ,\ \ n\geq 0
\end{equation}
For $n_0\geq 0$ and $\epsilon$ small, positive, define 
$\lambda_{n_0}=a_{n_0+2}\left(2+a_{n_0+2}\right)^{-1}-\epsilon$. Let
$\mathcal{N}_{n_0}$ be a small neighborhood of the interval
$I_{n_0}=[0,\lambda_{n_0}]$.  Consider the Banach space
$\mathcal{B}_{n_0}$ of sequences $\left\{\delta_n\right\}_{n\geq n_0}$
with $\delta_n=\delta_n(\lambda)$ analytic on
$\mathcal{N}_{n_0}$ and continuous up to the boundary, with the norm
$\|\delta\|=\sup_{n\geq n_0}\sup_{\lambda\in \mathcal{N}_{n_0}}
|\delta_n(\lambda)|$. Direct estimates show that the operator
$\mathcal{S}$ defined by (\ref{recdel}) takes 
the ball of radius $\rho_{n_0}=2/(2+a_{n_0+2})+\epsilon'$ in
$\mathcal{B}_{n_0}$ into itself (if $\epsilon,\epsilon'$ and
$\mathcal{N}_{n_0}$ are small enough), and is a contraction in this
ball. Therefore the equation $\delta=\mathcal{S}(\delta)$ has a unique
solution in $\mathcal{B}_{n_0}$, of norm less than $\rho_{n_0}$. Then
$\left|\delta_n(\lambda)\right|<\mbox{const}(n+2)^{-1/2}$ for all
$\lambda\in I_{n}$ and all $n\geq 0$. Since the sequence $\lambda_{n}$
increases to $1$, (i) follows.

(ii) {\em {Step I: all $g_n$ are meromorphic on $[0,1)$.}} 

Since $\delta_n$ is analytic on $I_n$, then from (\ref{dclr}),
$g_n$ is analytic on $I_n\setminus\{0\}$, having a pole at
$\lambda=0$: $g_n\sim \lambda^{-1}a_{n+1}(1+a_{n+1})^{-1}$
($\lambda\rightarrow 0$). Iterating (\ref{eq:contf}) it follows that
$g_{n-1},g_{n-2},...,g_0$ are meromorphic on $I_n$. Since the
intervals $I_n$ increase toward $[0,1)$ it follows that
$g_0,g_1,...g_n...$ are meromorphic on $[0,1)$.

{\em {Step II: there exists $n_1$ and $\lambda_0\in (0,1)$ such that
    $g_{n_1}(\lambda_0)\leq 0$.}}

Define $\epsilon_n=\left(1+a_n\right)^{-1}$; we have (see (\ref{an}))
\begin{equation}\label{epen}
\epsilon_{n_0}\sim r(n_0\omega)^{-1/2}\ \ ,\ \ n_0\rightarrow\infty
\end{equation}
Let $n_0$ be large and denote $\lambda_0=1-\epsilon_{n_0}$.  Let
$N_0$ be large enough so that $\lambda_0$ is in the domain of
analyticity of $g_{N_0}$. Iterating (\ref{eq:contf}) starting from $N_0$
(and decreasing indices) we get the value $g_{n_0}(\lambda_0)$. If for some
$n\in\{n_0,n_0+1,...,N_0\}$ we get $g_n(\lambda_0)\leq 0$, {\em {Step
    II}} is proved. Then assume that $g_{n_0}(\lambda_0)> 0$.

Consider the recurrence
\begin{equation}\label{rrtilde}
\tilde{r}_{n-1}=G_{n_0}(\lambda_0)-\frac{1}{\tilde{r}_n}\ \ \ {\mbox{for}}\ n\leq
n_0\ \ ,\ \tilde{r}_{n_0}=g_{n_0}(\lambda_0)
\end{equation}
where, in fact, $G_{n_0}(\lambda_0)=2-\epsilon_{n_0}^2$.

The recurrence (\ref{rrtilde}) can be solved
explicitly (it is a discrete Riccati equation and a substitution
$\tilde{r}_{n-1}=x_{n-1}/x_n$ transforms it into a linear recurrence
with constant coefficients).  It has the solution
\begin{equation}\label{solrntilde}
\tilde{r}_n=\frac{\cos\left((n-n_0)\phi+\theta\right)}{\cos\left((n+1-n_0)\phi+\theta\right)}
\end{equation}
where $ \cos\phi=1-{\epsilon_{n_0}^2}/{2}$, $\sin\phi>0$, and
$\tan\theta=(\cos\phi-\lambda)/\sin\phi$ so that
$\theta\sim\frac{\pi}{4}-\frac{1}{4}\epsilon_{n_0}$
($\epsilon_{n_0}\rightarrow 0$).

We assume, to get a contradiction, that $g_n(\lambda_0)>0$ for all
$n=0,1,...,n_1$. Then 
\begin{equation}\label{claim1}
g_n(\lambda_0)\leq\tilde{r}_n\ \ \ {\mbox{for}}\ n\leq n_0
\end{equation}
which follows immediately by induction using (\ref{eq:contf}),
(\ref{rrtilde}), noting that $G_n$ is increasing in $n$.

Note that there is an $n_1\in\{1,2,...,n_0-1\}$ so that 
\begin{equation}\label{claim2}
\tilde{r}_n>0 \ \ {\mbox{for}}\ \ n\in\{n_1+1,...,n_0\}\ \ \
{\mbox{and}}\ \ \tilde{r}_{n_1}<0
\end{equation}

Indeed (from (\ref{epen})) when $n$ decreases from $n_0$ the numerator
and denominator in (\ref{solrntilde}) increase up to 1, then decrease,
until the numerator becomes negative, when $n$ equals $n_1=n_0-k_1$
where $k_1$ is the integer with $k_1-1<(\pi/2+\theta)/\phi\leq k_1$.
Since $\phi\sim\epsilon_{n_0}$ ($\epsilon_{n_0}\rightarrow 0$) then
$k_1\sim(3\pi)/(4\epsilon_{n_0})$, and, using (\ref{epen}), clearly
$k_1\in \{1,...,n_0-1\}$ (if $n_0$ is sufficiently large).

Then (\ref{claim1}) and (\ref{claim2}) contradict the assumption that
$g_{n_1}(\lambda_0)>0$, and {\em{Step II}} is proved.

{\em{Step III.}} 

The function $g_{n_1}$ is meromorphic on $[0,1)$, with
$g_{n_1}(0+)=+\infty$. There is a smallest value of $\lambda$ in
$(0,\lambda_0)$ where $g_{n_1}$ changes sign: this is either a zero, or
a pole.

Assume it was a pole. Let $p\in(0,\lambda_0)$ be the first pole of
$g_{n_1}$. Then $g_{n_1}$ is positive and analytic on
$(0,p)$, and $g_{n_1}(p-)=+\infty$, $g_{n_1}(p+)=-\infty$. Since
$g_{n+1}=1/(G_{n+1}-g_n)$ (see (\ref{eq:contf})) then
$g_{n_1+1}(p-)=0-$, hence $g_{n_1+1}$ changes sign in $(0,p)$.
But $g_{n_1+1}$ has no zero in $(0,p)$ (otherwise at that zero
$g_{n_1}$ would have had a pole, from (\ref{eq:contf})). Then
$g_{n_1+1}$ has a pole, with a change of sign, from $+$ to $-$, in
$(0,p)$. Now the argument can be repeated. It follows that for any
$k>0$, $g_{n_1+k}$ has a pole in $(0,p)$, which contradicts the fact
that the domain of analyticity of $g_n$ increases to $(0,1)$ as
$n\rightarrow \infty$.

Therefore, the first change of sign of $g_{n_1}$ is at a zero. Let
$\zeta_1$ be the smallest value in $(0,1)$ such that $g_{n_1}(\zeta_1-)=0+$,
$g_{n_1}(\zeta_1+)=0-$. Then from (\ref{eq:contf}) we have
$g_{n_1-1}(\zeta_1-)=-\infty$ and $g_{n_1-1}$ changes sign in $(0,\zeta_1)$.
Now the argument can be repeated. It follows that $g_{0}$ has a pole
at a point $\zeta_{n_1}$ with $g_0(\zeta_{n_1}-)=-\infty$.
 
(iii) Since $g_0(\lambda)$ takes all the values when $\lambda\in
(0,\zeta_{n_1})$ there exists $\lambda=\lambda_s\in (0,1)$ such that
(\ref{inicd}) holds.

(iv) For $\lambda=\lambda_s$, since the solution of (i) satisfies
$g_n(\lambda)=\lambda^{-1}+O(n^{-1/2})$ we have from (\ref{VrV}),with
the notation $V^{[s]}=V(\lambda_s)$, that
$V_n^{[s]}=\prod_{k=0}^ng_k(\lambda_s)^{-1}V_0^{[s]}=O(\lambda_s^{n+o(n)})$
and thus $V_n^{[s]}-V_{n-1}^{[s]}=O(\lambda_s^{n+o(n)})$; then from
(\ref{eq:zV}) $z_n^{[s]}=O(\lambda_s^{n+o(n)})$.

(v) The substitution (variation of constants) $V_n=V_n^{[s]}v_n$
brings the recurrence (\ref{eq:V}) to a first order one: with the
notation $\Delta_n=v_n-v_{n-1}$ we have
$\Delta_n=V_{n-2}^{[s]}/V_{n}^{[s]}\Delta_{n-1}$ and the rest of the
argument consists of straightforward estimates.
\end{proof}

\subsubsection{Proof of Lemma \ref{homogsol}}

\begin{proof}
  
  Lemma \ref{contf}(v) shows that equation (\ref{eq:Zs}) has a unique
  (up to a multiplicative constant) small solution, $z_n^{[s]}\sim
  \lambda_s^{n+o(n)}$ ($n\rightarrow\infty$), while the general
  solution behaves like $z_n\sim \lambda_s^{-n+o(n)}$. Since $y_n\sim
  \sqrt{n}z_n$ the uniqueness of the $l_2$ solution is proven.

\end{proof}

\begin{figure}
\epsfig{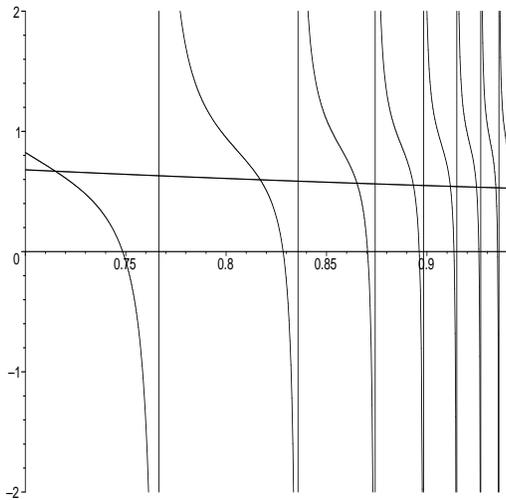}
\\
\caption{Graph of $g_0$ given by (\ref{eq:contf}) (discontinuous graph)
  and by (\ref{inicd}) in a region near $\lambda=1$, as functions of
  $\lambda$.  }
\end{figure}

\subsubsection{Examples of solutions}\label{exesol} 
We will show next how concrete values $\lambda_s $ satisfying
Lemma~\ref{contf}~(iii) are relatively straightforwardly, and
rigorously, found. One method is as follows. Note that the
minimum/maximum of the function $a-b/x$ where $x$ varies in an
interval not containing zero is achieved at the endpoints.  We thus
take the recurrence (\ref{eq:contf}) with initial conditions
$g_{n_0}=\lambda^{-1}\pm \frac{1-\lambda^2}{\sqrt{n\omega}}$ and
compute $g_0$ from these. The actual graph will be between these two,
unless the condition mentioned is violated in between $n_0$ and $0$.
This graph is to be intersected with the graph of the initial
condition (\ref{inicd}).

We take for instance $\omega=1.1$, $r=0.45$, $s_p=0.11$, $n_0=10$, for
which the rigorous control is not too involved. The two graphs are
very close to each other (within about $3.10^{-6}$ for
$\lambda\in(0.3,0.4)$) and cannot be distinguished from each-other in
Figure 1.  A first intersection is seen at $\lambda\approx 0.327$; see
Figure 2.

\begin{figure}
\epsfig{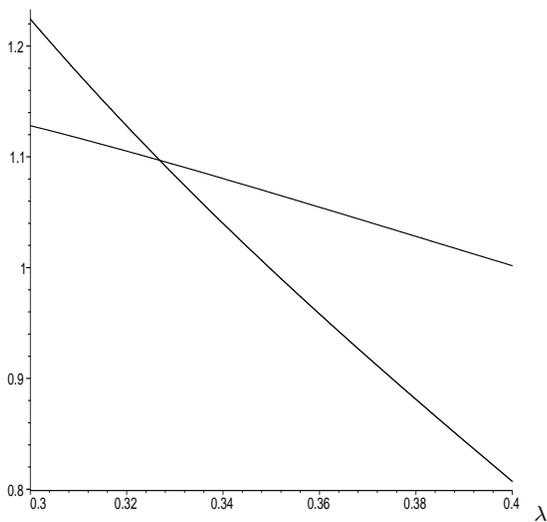}\vskip 0.5cm
\caption{Graphs of $g_0$ (steeper graph) and of \hskip -4ex $\
^{^{^{^{^{^{\displaystyle \lambda}}}}}}$   the initial
  condition
for $g_0$ (\ref{inicd}).}
\end{figure}

\subsection{Proof of Lemma \ref{keradj}}

Denote $B=(I-{\mathcal{J}}_0)A$;  we have  $B=A-S$.
Hence $B^*=\overline{A}-S$. Then $\mbox{Ker}(B)=\mbox{Ker}(B^*)$
(since $Az=Sz$ implies (\ref{whichzn0}), so $Az=\overline{A}z$, and
similarly, $\overline{A}z=z$ implies $Az=\overline{A}z$). So
$\mbox{Ker}[(1-{\mathcal{J}_0})A]=\mbox{Ker}[A(1-{\mathcal{J}_0}^*)]$
so that (since $A$ is one-to-one)
$A^{-1}\mbox{Ker}(1-{\mathcal{J}_0})=\mbox{Ker}(1-{\mathcal{J}_0}^*)$,
which proves the Lemma.

\subsection{Discussion of the singularities of solutions of (\ref{eq:rec3})}

Let $\lambda=\lambda_s$.  We have that $I-\mathcal{J}$ is invertible
for $\Re p_0>0$, and is not invertible at $p_0=is_p$ (Lemma
\ref{homogsol}). By the analytic Fredholm theorem (see e.g.
\cite{Reed-Simon}) $(I-\mathcal{J})^{-1}$ is meromorphic on a small
neighborhood of $is_p$, therefore there exist $m\geq 1$ and operators
$S_m,...,S_1,R(p_0)$ so that:
\begin{multline}\label{meroinv}
\left( I-\mathcal{J}\right)^{-1}=\frac{1}{\left(
    p_0-is_p\right)^m} S_{m}+...+\frac{1}{ p_0-is_p} S_{1}+R(p_0)
\end{multline}
where $R(p_0)$ is analytic at $is_p$, and $S_{m}\ne 0$ (since
$I-\mathcal{J}_0$ is not invertible).  Multiplying (\ref{meroinv}) by
$I-\mathcal{J}$ to the left, respectively to the right, and writing
$\mathcal{J}=\mathcal{J}_0+(p_0-is_p)\mathcal{J}_1(p_0)$ (where
$\mathcal{J}_1(p_0)$ is analytic at $is_p$) we get that
$$R_1(p_0)=\frac{1}{\left( p_0-is_p\right)^m} \left(
  I-\mathcal{J}_0\right) S_{m}+O\left( \left(
    p_0-is_p\right)^{-m+1}\right)$$
$$R_2(p_0)=\frac{1}{\left( p_0-is_p\right)^m}S_{m}\left(
  I-\mathcal{J}_0\right) +O\left( \left(
    p_0-is_p\right)^{-m+1}\right)$$
where $R_{1,2}$ are analytic at
$p_0=is_p$. By the uniqueness of the series of the analytic functions
(Banach space valued) $R_{1,2}$ we must then have
\begin{equation}\label{orel12}
\left( I-\mathcal{J}_0\right) S_{m}=0=S_{m}\left(
  I-\mathcal{J}_0\right)
\end{equation}

The first equality in (\ref{orel12}) implies
$\mbox{Ran}(S_{m})\subset \mbox{Ker}\left(
  I-\mathcal{J}_0\right)=\bigvee \{ y_{Ker}\}$ and since $S_{m}\ne 0$
then $\mbox{Ran}(S_{m})=\bigvee \{ y_{Ker}\}$ therefore
$S_{m}y=\langle y,u \rangle y_{Ker}$ for some $u\in l_2\setminus\{0\}$. The second
equality in (\ref{orel12}) means $u\in \mbox{Ran}\left(
  I-\mathcal{J}_0\right)^\perp={\mbox{Ker}}\left(
  I-\mathcal{J}_0^*\right)$.

By Lemma \ref{keradj} then (up to a multiplicative constant)
$u=A^{-1}y_{Ker}=z_{Ker}$ where $z_{Ker}$ satisfies (\ref{recuzngen}),
hence (\ref{eq:Zs}),(\ref{eq:Zsini}). The solution $y=\left(
I-\mathcal{J}\right)^{-1}f$ of (\ref{eq:rec3}) is then singular at
$p_0=is_p$ if $c=<f,z_{Ker}>\ne 0$.  For the example of \S\ref{exesol}
this latter condition can be checked  by explicit calculation
of the truncations to 10 terms and estimation of the remainder based on
the contractivity bounds in the previous section.  The result is
$c=-1.953\pm 0.001$. Thus the inhomogeneous equation has poles.

\begin{Lemma}\label{lnicep}

Let $Y(t)$ be analytic on $[0,\infty)$, with
$\lim_{t\rightarrow\infty}Y(t)=0$.

Let $s\in\RR$. Then 
\begin{equation}\label{nicep}
\lim_{a\downarrow 0} a\int_0^\infty e^{-(a+is)t}Y(t)\, dt=0
\end{equation}

\end{Lemma}

\begin{Corollary}
  
  Let $Y(t)$ be as in Lemma \ref{lnicep}. Let $y(p)=\int_0^\infty
  e^{-pt}Y(t)\, dt$. Assume that $y(p)$ is analytic on $i\RR_+$,
  except for a set of isolated points. Then $y(p)$ does not have poles
  on $i\RR_+$.

\end{Corollary}

\begin{proof}

{\em{I.}} We first show (\ref{nicep}) for $s=0$.

Separating the positive and negative parts of $\Re Y(t)$, $\Im Y(t)$
write $Y(t)=Y^{[1]}(t)-Y^{[2]}(t)+iY^{[3]}(t)-iY^{[4]}(t)$ with $Y^{[k]}(t)$ nonnegative,
continuous, nonanalytic only on a discrete set, where the left and right
derivatives exist, with $\lim_{t\rightarrow\infty}Y^{[k]}(t)=0$. It is
enough to show (\ref{nicep}) for each $Y^{[k]}$. 
Let then $Y$ be one of the $Y^{[k]}$'s. Denote $H(t)=\sup_{\tau\geq
  t}Y(\tau)$. The function $H$ on $[0,\infty)$ has the same properties
as $Y$ and in addition is decreasing. Then $H'$ exists a.e. and $H'\in
L^1[0,\infty)$ since $\int_0^\infty
  |H'(\tau)|\,d\tau=-\lim_{t\rightarrow\infty}\int_0^t
  H'(\tau)\,d\tau= \lim_{t\rightarrow\infty}-H(t)+H(0)=H(0)$

Then 
$$a\int_0^\infty e^{-at}Y(t)\, dt\leq   a\int_0^\infty e^{-at}H(t)\,
dt=-\int_0^\infty \frac{d}{dt}\left(e^{-at}\right)H(t)\, dt$$
$$=H(0)+\int_0^\infty e^{-at}H'(t)\, dt$$
therefore
$$\lim_{a\downarrow 0} a\int_0^\infty e^{-at}Y(t)\, dt\leq
H(0)+\lim_{a\downarrow 0}\int_0^\infty e^{-at}H'(t)\, dt=0$$
which proves the Lemma in this case.

{\em{II.}} Let now $s\in\RR$ arbitrary. Then (\ref{nicep}) follows
from the result for $s=0$ applied to the function
$\tilde{Y}(t)=e^{-ist}Y(t)$.

\end{proof}

{\bf{Proof of Theorem \ref{T2}}}

In conclusion $Y(t)$ cannot tend to zero as $t\rightarrow\infty$ and
complete ionization fails.

\textbf{Acknowledgments.} The authors would like to thank A. Soffer and
M. Weinstein  for interesting discussions and
suggestions.  Work of O. C. was supported by NSF Grant 9704968, of
R.D.C. by NSF Grant 0074924, and that of O. C., 
J.  L. L. and A. R. by AFOSR Grant F49620-98-1-0207 and NSF Grant
DMR-9813268.

\section*{Appendix A. Proof of Lemma~\ref{expEstim}}

\makeatletter
\@addtoreset{equation}{section}
\makeatother

\setcounter{section}{1}
\renewcommand{\theequation}{\Alph{section}\arabic{equation}}
\renewcommand{\thesubsection}{\thesection \alph{subsection}\ }

\setcounter{equation}{0}

(i)
  Consider $L^1_{loc}[0,A]$ endowed with the norm $\|F\|_{\nu}:=\int_0^A
  |F(s)|e^{-\nu s}ds$, where $\nu>0$. If $f$ is continuous and
  $F,G\in L^1_{loc}[0,A]$, a straightforward calculation shows that

  \begin{align}
    \label{eq:properties}
\|fF\|_{\nu}<\|F\|_\nu\sup_{[0,A]}|f|
\\
\|F*G\|_{\nu}<\|F\|_\nu\|G\|_\nu
\\
\|F\|_\nu\rightarrow 0\ \mbox{ as }\ \nu\rightarrow\infty
  \end{align}
  
  \z where the last relation follows from the Riemann-Lebesgue lemma.

\z The integral equation (\ref{eq:(5)}) can be written as 

\begin{equation}
  \label{eq:eqa}
  Y=\eta+\mathcal{J}Y \mbox{ where } \ \ \ \mathcal{J} F:= \eta(2i+M)*F
\end{equation}
  Since $M$ is locally in $L^1$ and bounded for large $x$ it is clear
  that for large enough $B_2$, (\ref{eq:(5)}) is contractive if
  $\nu>B_2$, for any $A$.

(ii)
 This is an immediate consequence of Lemma~\ref{expEstim} and of
 elementary properties of the Laplace transform.

(iii) We have in $\mathbb{H}$,

\begin{align}\label{intres}
  \mathcal{L}M=\lim_{a \downarrow 0}\frac{2i}{\pi} \int_0^{\infty}\drm x
  e^{-px}\int_0^{\infty}\frac{u^2e^{-i(x-ia)(1+u^2)}}{1+u^2}\drm
  u\\=\frac{i}{\pi}\int_{-\infty}^{\infty}\frac{u^2}{(1+u^2)(p+i(1+u^2))}\drm u
\end{align}

 For $\Re(p)>0$ we push the
integration contour through the upper half plane. At the two poles in
the upper half plane $u^2+1$ equals $0$ and $ip$ respectively, so that

\begin{multline}
 \frac{i}{\pi} \int_{-\infty}^{\infty}\frac{u^2}{(1+u^2)(p+i(1+u^2))}\drm
  u\\=\frac{i}{\pi}
\left(
\frac{(-1)}{(2i)(p)}\oint\frac{ds}{s}+\frac{u_0^2}{(ip) (2
  iu_0)}\oint\frac{ds}{s}\right)
=-\frac{i}{p}+\frac{u_0}{p}
\end{multline}

\z where $u_0$ is the root of $p+i(1+u^2)=0$
in the {\em upper} half plane. Thus

\begin{equation}
  \label{eq:e23}
 \mathcal{L}M =
-\frac{i}{p}+\frac{i\sqrt{1-ip}}{p}
\end{equation}

\z with the branch satisfying $\sqrt{1-ip}\rightarrow 1$ as $p\rightarrow 0$ in
$\mathbb{H}$. As $p$ varies in $\mathbb{H}$, $1-ip$ belongs to the lower
half plane $-i\mathbb{H}$ and then $\sqrt{1-ip}$ varies in the fourth
quadrant.

For $\Re(p)>0,\, \omega>0$ we have

\begin{align}
  \mathcal{L}\Big(e^{\pm i\omega}M\Big)=-\frac{i}{p\mp
  i\omega}+\frac{i\sqrt{1-ip\mp\omega}}{p\mp i\omega}\\
(\mbox{with }\ \ \ \sqrt{1-ip-\omega}=-i\sqrt{\omega-1+ip}\ \mbox{ if }\ \omega>1)\nonumber
\end{align}

\z and relation (\ref{fnceq}) follows.

\section*{Appendix B.  Discussion of the genericity condition (g)}

\setcounter{section}{2}
\setcounter{equation}{0}
 A thorough analysis
of the properties of the shift operator is provided by the treatise
\cite{Nikol'skii}. We provide here an independent discussion, meant to
give an insight on the
interesting analytic properties involved in this condition.

Let ${C}=(C_0,C_1,...,C_n,...)\in l_2(\NN)$ and the operator
${T}$ defined as before by ${T}{C}=(C_1,C_2,...)$. We want to see
for which such vectors, the system of equations

\begin{equation}
  \label{eq:translates}
  (\bfz,{T}^j\bfC)=0,\ j=0,1,...
\end{equation}

\z has nontrivial solutions $\bfz$ in $l_2$. We can associate to 
$\bfz$ and $\bfC$ analytic functions in the unit disk, $Z(x)$ and $C(x)$
 by

 \begin{equation}
   \label{eq:an}
   C(x)=\sum_{k=0}^{\infty}C_kx^k\ \ \ Z(x)=\sum_{k=0}^{\infty}z_kx^k\
 \end{equation}

\z These functions, extend to $L^2$ functions on the unit circle. The
system
of equations (\ref{eq:translates}) can be written as

\begin{multline}
  \label{eq:an1}
  z_0C(x)+z_1
  x^{-1}(C(x)\-C(0))+...\\+z_n\left[x^{-n}C(x)-x^{-n}\sum_{k=0}^{n-1}\frac{x^k}{k!}C^{(k)}(0)\right]+...=0
\end{multline}

\z Using Cauchy's formula, we can the difference in square brackets
in(\ref{eq:an1}) as

\begin{equation}
  \label{eq:Cauchy1}
  \frac{1}{2\pi i}\oint_{|s|=1}\frac{C(s)}{s^n(s-x)}ds
\end{equation}

\z and thus (\ref{eq:translates}) becomes

\begin{equation}
  \label{eq:aneq}
  \oint_{|s|=1}\frac{C(s)Z(1/s)}{s-x}ds=0
\end{equation}

\z The functions $C$ for which this equation has nontrivial solutions
$Z$ relate to the Beurling inner functions \cite{Nikol'skii} and 
are very ``rare''.

{\bf Examples}. (i) Let $|\lambda|<1$ and $C_n=\lambda^n$, i.e.
$C(x)=(1-\lambda x)^{-1}$. This is related to the function
(\ref{eq:counterex}).  Taking advantage of the analyticity of $Z$
outside the unit circle, we can push the contour of integration towards
$s=\infty$, collecting the residue at $x=\lambda^{-1}$; we see that
equation (\ref{eq:aneq}) holds iff $Z(\lambda)=0$, i.e., for a space of
analytic functions of codimension one.

(ii) Let $\lambda_n=1/n$. Then $C(x)=\ln(1-x)$, and by taking
$s=1/t$ in (\ref{eq:aneq}) we get

\begin{equation}
  \label{eq:ln}
  \frac{1}{x}\oint_{|t|=1}\frac{Z(t)\ln(t-1)}{(t-x^{-1})t}dt
-\frac{1}{x}\oint_{|t|=1}\frac{\ln(t)Z(t)}{t(t-x^{-1})}dt=0
\end{equation}

\z By making a cut on $[1,\infty)$ for the $\log$ we see that the
integrand
in the first
integral is analytic in the unit circle and thus the integral vanishes.
We decompose the second integral by partial fractions and we get

\begin{equation}
  \label{eq:ln2}
\oint_{|t|=1}\frac{\ln(t)Z(t)}{t}dt-\oint_{|t|=1}\frac{\ln(t)Z(t)}{(t-y)}dt=0
\end{equation}

\z where $y=x^{-1}$. The first integral is a constant, $C$. By pushing
the contour of integration inwards, we see that the second integral
extends analytically for small $y\ne 0$. For such $y$ we thus have

\begin{equation}
  \label{eq:an3}
  \oint_{|t|=1}\frac{\ln(t)(Z(t)-Z(y))}{(t-y)}dt+Z(y)\oint_{|t|=1}\frac{\ln(t)}{(t-y)}dt=-C
\end{equation}

\z Now the contour of integration can be pushed to the sides of the
interval $[0,1]$ collecting the difference between the branches of the
log.
We get

\begin{equation}
  \label{eq:an4}
  \int_0^1\frac{Z(t)-Z(y)}{t-y}dt+Z(y)\int_0^1\frac{1}{t-y}dt=0
\end{equation}

\z Thus $\phi(y)+Z(y)\ln(-y)=C$ with $\phi$ and $Z$ analytic in the unit
circle
thus $\ln(-y)$ is analytic unless $Z=0$. This shows $C_n=1/n$ is
generic.

\end{document}